\documentclass[twocolumn,amsmath,amssymb]{revtex4}
\usepackage{graphicx}
\usepackage{epsfig}

\begin{document}
\title{Ballistic Hot Electron Transport in Graphene}

\author{Wang-Kong Tse}
\author{E. H. Hwang}
\author{S. Das Sarma}
\affiliation{Condensed Matter Theory Center, Department of Physics,
University of Maryland, College Park, Maryland  
20742-4111, USA}


\begin{abstract}
We theoretically study the inelastic scattering rate and the carrier mean free path for energetic hot electrons in graphene, including both electron-electron and electron-phonon interactions. Taking account of optical phonon emission and electron-electron scattering, we find that the inelastic scattering time $\tau \sim 10^{-2}-10^{-1}\,\mathrm{ps}$ and the mean free path $l \sim 10-10^2\,\mathrm{nm}$ for electron densities $n  = 10^{12}-10^{13}\,\mathrm{cm}^{-2}$. In particular, we find that the mean free path exhibits a finite jump at the phonon energy $200\,\mathrm{meV}$ due to electron-phonon interaction. Our results are directly applicable to device structures where ballistic transport is relevant with inelastic scattering dominating over elastic scattering. 
\end{abstract}

\maketitle




The existence \cite{SSC} of gated two-dimensional (2D) graphene layers, where carrier 
transport controlled by an external gate has become possible \cite{N1}, provides the exciting 
possibility of novel high-speed electronic device structures \cite{app} utilizing the high 
graphene carrier mobility \cite{N1,N2}. Such fast graphene devices would work in the ballistic transport 
regime, where carrier mobility limited by elastic scattering, is essentially irrelevant (i.e. $l_{\mathrm{e}} > l$, with  $l_{\mathrm{e}}$, $l,$ being respectively the elastic and the inelastic carrier mean free path), and what matters is the 
inelastic scattering due to electron-electron and electron-phonon interactions. Such ballistic devices for ultrafast applications can only work if the relevant device dimensions are smaller than the inelastic mean free path $l$, and the speed of this device is limited by the inelastic scattering time $\tau$ ($< \tau_{\mathrm{e}}$, where $\tau_{\mathrm{e}}$ is the elastic relaxation time). In currently available high-mobility ($> 20,000\,\mathrm{cm}^2/\mathrm{Vs}$) graphene samples, $l_{\mathrm{e}} (\tau_{\mathrm{e}}) \gtrsim 10^3\,\mathrm{nm}\,(1\,\mathrm{ps})$, and therefore, inelastic scattering will dominate device operations for length (time) scales below $10^3\,\mathrm{nm}\,(1\,\mathrm{ps})$. 

In this Letter, we calculate the inelastic mean free path ($l$) and the corresponding inelastic scattering rate $\tau^{-1}$ in graphene limited by electron-electron and electron-phonon interactions. Our study is motivated by
electron transport in the hot electron transistor device structure
which is so designed as to allow electrons to traverse
the base region ballistically.  
In such a device scheme, highly energetic electrons are injected in
the emitter region which then travel through the base region
ballistically before reaching the collector region. 
The fraction of electrons $\alpha$ that reach the collector goes as
$\alpha \sim e^{-d/l}$, where $d$ is the width of the base region. 
The mean free path is given by $l = v\tau$, 
where $v$ is the Fermi velocity of electron and $\tau$ is the inelastic 
scattering time, which, in general, is a strong function of the
injected energy of electrons and the electron density in the base
region. 

We consider two principal mechanisms contributing to inelastic
scattering arising from many-body interactions: (1) absorption or
emission of optical phonons by electrons due to electron-phonon (e-ph)
interaction; (2) exchange-correlation effects induced by
electron-electron (e-e) interaction. 
For the e-ph interaction, we take into account
the most dominant phonon mode in graphene -- the LO phonons at the
Brillouin zone center $\Gamma$. This mode shows up as the `G peak'
resonance observed in Raman scattering experiments with a phonon
energy $\omega_0 \approx 200\,\mathrm{meV}$ \cite{Raman}.  
Acoustic phonons couple very weakly to electrons in graphene, and the
associated scattering rates are on the order of $10^{11}$ s$^{-1}$
even at room temperature 
\cite{chen_phonon} and can be ignored when compared to the scattering mecanism discussed
above. We also neglect electron-impurity scattering in our calculation
because the mean free paths of currently available high mobility 
graphene samples are much longer than the inelastic mean free paths to
be calculated in this paper.


The inelastic scattering of electrons causes damping (i.e., decay) of
the quasiparticle state, and the inelastic scattering lifetime
$\tau(k)$ is given by the imaginary part of the self-energy
$\mathrm{Im}\Sigma(k,E)$ evaluated on the energy shell
$E = \xi_k$, where $\xi_k = \varepsilon_k-\mu$ ($\varepsilon_k = \hbar vk$ is the electron kinetic energy 
and $v \approx 10^6\,\mathrm{ms}^{-1}$ is the Fermi velocity which is constant 
in graphene irrespective of electron density) is the single-particle 
energy rendered from the chemical potential $\mu$. The inelastic
scattering rate (or damping rate) $1/\tau$ consists of two
contributions, given by  
%
${\hbar}/{2\tau(k)} =
\mathrm{Im}\Sigma_{\mathrm{e-ph}}^{\mathrm{R}}(k,\xi_k)+\mathrm{Im}
\Sigma_{\mathrm{e-e}}^{\mathrm{R}}(k,\xi_k)$,   
%
where the first term denotes the contribution to the self-energy due 
to electron-LO phonon interaction \cite{tsephonon}, 
\begin{eqnarray}
&&\mathrm{Im}\Sigma_{\mathrm{e-ph}}^{\mathrm{R}}(k,\xi_k) =
-{\pi}g^2\sum_{\lambda = \pm
  1}\sum_{k'}\left\{n_F(\xi_{k'\lambda})\right. \nonumber \\ 
&&\left.\delta(\xi_k-\xi_{k'\lambda}+\omega_0) +
  [1-n_F(\xi_{k'\lambda})]\delta(\xi_k-\xi_{k'\lambda}-\omega_0)\right\}  
\nonumber \\ 
&&\frac{1-\lambda\mathrm{cos}(\phi_{k'}-2\phi_{k'-k})}{2}, \label{eq2}
\end{eqnarray}
and the second term on the right denotes the contribution to the
self-energy due to e-e Coulomb interaction
\cite{dassarma_ee}, 
\begin{eqnarray}
&&\mathrm{Im}\Sigma_{\mathrm{e-e}}^{\mathrm{R}}(k,\xi_k) = \label{eq3} \\
&&\sum_{\lambda = \pm
  1}\sum_{k'}[n_B(\xi_{k'\lambda}-\xi_k) +
n_F(\xi_{k'\lambda})]V_{\mathbf{k}'-\mathbf{k}} \nonumber \\  
&&\mathrm{Im}\left[\frac{1}{\epsilon(\mathbf{k}' -
    \mathbf{k},\xi_{k'\lambda}-\xi_k+i0^+)}\right]\frac{1 +
  \lambda\mathrm{cos}(\phi_{k'}-\phi_k)}{2}.  
\nonumber  
\end{eqnarray}
Here, $n_B(x), n_F(x) = 1/(\mathrm{exp}(x/k_{\mathrm{B}}T)\mp 1$ are
the Bose and Fermi distribution functions, respectively. For the
e-ph contribution Eq.~(\ref{eq2}), $g = -(\beta\hbar
v/a^2)\sqrt(\hbar/2NM_{\mathrm{c}}\omega_0)$ is the e-ph 
coupling constant \cite{Ando}, where $\beta \simeq 2$ is a
dimensionless constant characterizing the rate of change of the
nearest-neighbor hopping energy with respect to bond length, $a = 1.42
\,\mathrm{\AA}$ is the equilibrium bond length, $N$ is the number of
unit cells, $M_{\mathrm{c}} = 2.2 \times 10^4\,m_{\mathrm{e}}$ is the
mass of a carbon atom ($m_{\mathrm{e}}$ is the electron mass). For the
Coulomb contribution Eq.~(\ref{eq3}), $V_q = 2\pi e^2/q$ is the bare
Coulomb interaction, $\epsilon(q,\omega) = 1-V_q\Pi(q,\omega)$ is the
dielectric function given, within the random phase approximation, by
the electron polarizability $\Pi(q,\omega)$ \cite{Polar}. 
%
\begin{figure}[h]
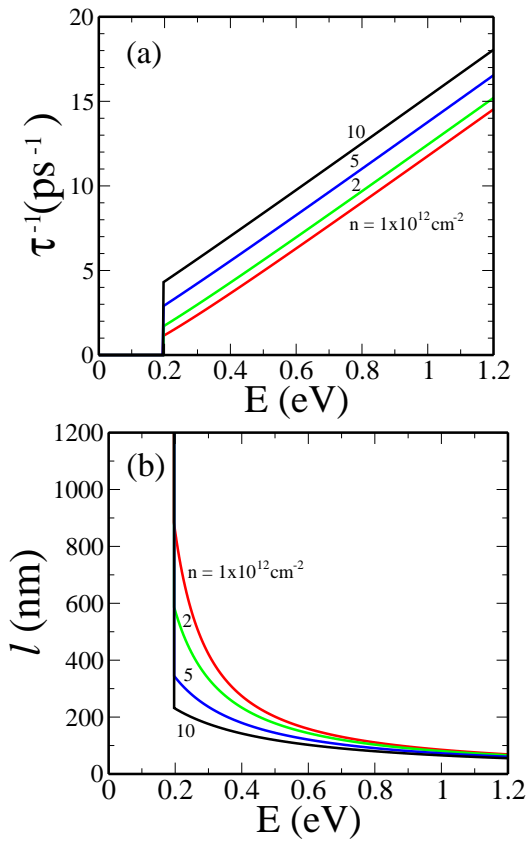

  \includegraphics[width=6.5cm,angle=0]{fig_ep_a.eps} 
  \includegraphics[width=6.9cm,angle=0]{fig_ep_b.eps} 
\caption{(a) Inelastic scattering rate due to e-ph 
  interaction versus the single-particle energy of hot electron $E = \xi_k$ 
  for different electron densities $n$. $n = 1, 2, 5, 10 \times
  10^{12}\mathrm{cm}^{-2}$ are indicated by the red, green, blue and
  black lines, respectively. (b) Corresponding inelastic mean free
  path $l$.} 
\label{fig1}
\end{figure}

In the following, we calculate $1/\tau(k)$ at zero temperature, which
is a very good approximation for graphene since at the usual doping
density $n = 10^{12} - 10^{13} \mathrm{cm}^{-2}$, the corresponding
Fermi temperature $T_F \simeq 1400 - 4300 \mathrm{K}$ is much higher
than room temperature. First we calculate the self-energy correction
due to e-ph interaction, Eq.~(\ref{eq2}).  
%
%
Fig.~\ref{fig1}(a) shows the scattering rate $1/\tau$ versus electron energy $E = \xi_k$ for different 
values of electron density. The gap from $0$ to $0.2\,\mathrm{eV}$
(which is the LO phonon energy $\omega_0$) is a characteristic feature of the LO 
phonon absorption process; it results from the Pauli blocking by those 
electrons located within an amount of energy $\omega_0$ of the Fermi 
level, so that decay by electrons with energy $\xi_k
\in [-\omega_0,\omega_0]$ is forbidden due to the restricted phase space. 
Beyond the gap, $1/\tau$ behaves linearly as $k$ due to the linear 
dependence on momentum of the graphene density of states,
$\nu(k) = 2k/\pi \hbar v$. As a result of the gap in $1/\tau$, the calculated mean free
path $l = v\tau$ [Fig.~\ref{fig1}(b)] is infinite within the range of the
phonon energies, before falling off as $\sim 1/E$ with electron energy 
$E$. Fig.~\ref{fig1} also shows that the scattering rate (mean free path)
increases (decreases) with electron density, as scattering events
become more frequent with increasing number of electrons.  


\begin{figure}[h]
  \includegraphics[width=6.3cm]{fig_ee_a.eps}
  \includegraphics[width=6.3cm]{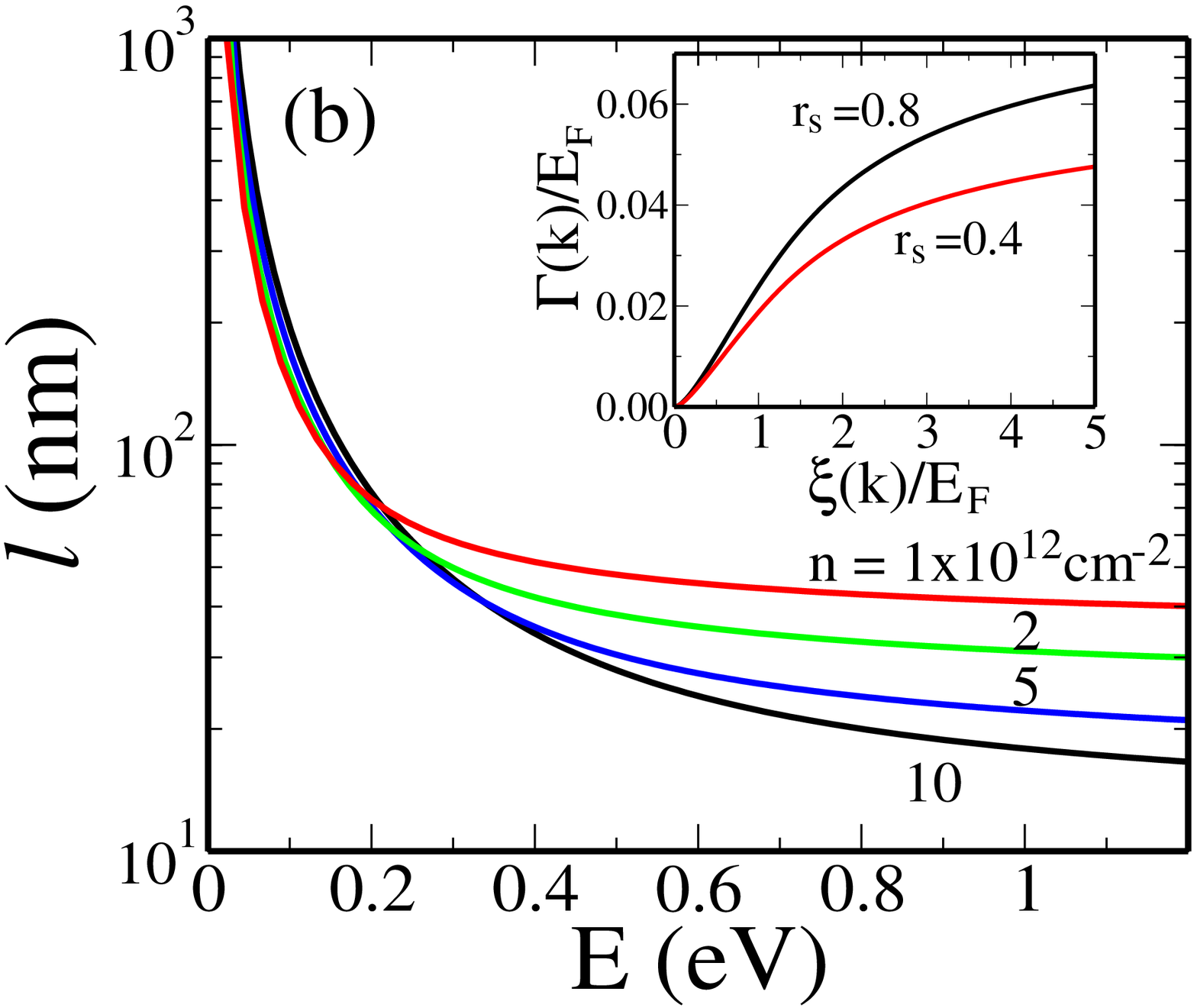}
\caption{Calculated scattering rate (a) and the corresponding mean free
  path (b) of hot electron as a function of energy $E =
  \xi_k$ for different carrier densities $n=1$, 2, 5, 10$\times 10^{12}$ cm$^{-2}$. 
The inset in (b) shows the calculated 
  damping rate $\Gamma(k) \equiv \hbar/2\tau(k)$ scaled by Fermi energy as a function of energy
  divided by Fermi energy for different coupling constant $r_s=0.8$
  and $0.4$, which correspond to graphene on SiO$_2$ and SiC substrates, respectively. 
Note the scaled damping rates are independent on the density.} 
\label{fig_ee}
\end{figure}

For e-e Coulomb interaction, we show the calculated 
inelastic scattering rate Eq.~(\ref{eq3}) and the 
corresponding inelastic mean free path, $l = v\tau$ in Fig.~\ref{fig_ee}. 
The strength of Coulomb interaction is characterized by the
dimensionless coupling parameter   
 $r_s =e^2 /\kappa \hbar v$, where $\kappa$ is the effective background
 dielectric constant of the substrate \cite{Polar}. In the inset of
 Fig. \ref{fig_ee}(b), the calculated damping rates are shown 
for different coupling constants $r_s =0.8$ and $r_s=0.4$. 
Before discussing the scattering time we consider the scattering rate $1/\tau$ 
due to the e-e interaction.
In conventional parabolic-band semiconductors, an electron 
injected with sufficient kinetic energy can decay via both plasmon
emissions and single-particle intraband excitations (i.e. Landau
damping) \cite{jalabert}. In doped graphene, however, injected 
electrons cannot decay via plasmon emission due to phase space
restrictions \cite{hwang_ee}.  
Multiparticle excitations, which are excluded in the approximations
used here, will constitute finite damping of the quasiparticles, but the effects of such higher-order processes 
are relatively small in graphene. (Note that in undoped graphene even 
single-particle excitations are forbidden, so that the scattering rate 
within the Born approximation is zero due to electron-electron Coulomb
interaction at $T=0$ \cite{dassarma_ee}.) 
Since only single-particle excitations give rise to damping of 
the quasiparticles, for electron energy close to the Fermi level the calculated scattering rate 
due to e-e interaction in graphene, similar to the case of 2D parabolic-band 
semiconductors, is given by $1/\tau \sim |\varepsilon_k- E_F|^2 \mathrm{ln}
|\varepsilon_k -  E_F|$ \cite{dassarma_ee,hwang_ee}. 
Farther away from $E_F$, however, the dependences of
$1/\tau(k)$ on $k$ in graphene and in
parabolic-band semiconductors are  
\textit{qualitatively} different because both plasmon emissions and interband
processes are absent in graphene. On the other hand, the only independent 
parameters relevant for the quasiparticle scattering 
rate under the Born approximation at $T=0$ are the Fermi energy 
and the dimensionless coupling constant $r_s$. Therefore the
calculated scattering rate due to  
e-e interaction, which has units of energy, must be 
proportional to $E_F$, and must be a function only of $\varepsilon_k/E_F =
k/k_F$ and $r_s$ [see the inset of Fig. \ref{fig_ee}(b)]. 

In Fig. \ref{fig_ee}(a) the inverse lifetime in units of inverse of ps is shown
as a function of energy $E = \xi_k$  
of the hot electrons. Below $200$ meV, which corresponds to the optical phonon 
energy of graphene, the calculated scattering time is only weakly
dependent on the carrier density of the system. However, for very 
energetic hot electron $E \agt 1$eV the scattering time shows a 
strong density dependence and increases by almost a 
factor of two as the density decreases from $n=10^{13}$ cm$^{-2}$ to 
$n=10^{12}$ cm$^{-2}$. Our results are consistent with recent 
experiments of decay time in ultrafast carrier \cite{dawlaty}, 
in which the measured decay times are in the $0.07-0.12$ ps
range. The corresponding inelastic mean free path is 
shown in Fig. \ref{fig_ee}(b) as a function of energy. We find that the
characteristic mean free path of the hot electron with $100$ meV energy
above the Fermi energy is about $100$ nm. Again below $200$ meV the
calculated mean free path is weakly dependent on density. 

Combining the results from the e-ph (Fig.~1) and e-e interactions (Fig.~2), the 
total scattering rate becomes non-zero below the phonon energy $200\,\mathrm{meV}$, 
and the mean free path $l \sim 10-10^2\,\mathrm{nm}$ for injected energy $E < 200\,\mathrm{meV}$ whereas 
$l \sim 10\,\mathrm{nm}$ for $E > 200\,\mathrm{meV}$.  

We now propose an interesting device principle based on our results in
this paper, which are peculiar to the many-body effects in graphene. In doped (or
gated) graphene, the dominant inelastic scattering process of hot electrons 
below $200$ meV comes from intraband single-particle excitation due to
screened electron-electron interaction; above $200$ meV, 
inelastic scattering due to e-ph interaction sets in as electrons are now able to
emit LO phonons. A lateral hot-electron transistor (LHET) device \cite{ltheta} 
where the electrons travel in the graphene sheet of the base region
can in principle be fabricated, whose operation makes use of the abrupt change in the inelastic mean 
free path due to electron-coupled mode scatterings of the injected electrons. 
Thus, by varying the inelastic scattering rate through changing the injection energy, one 
can achieve a significant change in the electron mean free path and hence 
the emitter-collector current. It is also possible to use the peculiar scattering rate of undoped (or ungated) 
graphene, which is totally suppressed due to phase space restrictions 
\cite{dassarma_ee}. With the application of gate voltage one can easily tune the Fermi 
level through the electron density, so that the damping process via e-e or/and 
e-ph interaction can be activated and deactivated. 

In conclusion, we have calculated the inelastic scattering rate $1/\tau$ and the inelastic mean free path $l$ in graphene. We find that $\tau \sim 10^{-2}-10^{-1}\,\mathrm{ps}$ and $l \sim 10-10^2\,\mathrm{nm}$, with a finite step jump at $200\,\mathrm{meV}$ at the LO phonon energy. Our results have direct relevance to ballistic transport in graphene fast device structures. 

This work is supported by US-ONR, NRI-SWAN and NSF-NRI.


\begin{thebibliography}{18}

\bibitem{SSC} See, for example, the Graphene special issue of Solid
State Communications, vol.\textbf{143}, p.1-123 (2007), edited by 
S. Das Sarma, A.K. Geim, P. Kim, and A.H. MacDonald. 


\bibitem{N1} K. S. Novoselov \textit{et al}., Science {\bf 306}, 666 (2004); Y.-W. Tan \textit{et al}., Phys. Rev. Lett. {\bf 99} 246803 (2007); J.-H. Chen \textit{et al}., Nat. Phys. {\bf 4}, 377 (2008). 


\bibitem{app} M.C. Lemme \textit{et al}., IEEE Electron Device Lett. {\bf 28}, 282 (2007); 
J.R. Williams, L. DiCarlo and C.M. Marcus, Science {\bf 317}, 638 
(2007); G. Liang \textit{et al}.,
IEEE Trans. Electron Devices {\bf 54}, 657 (2007);
G. Gu \textit{et al}., Appl. Phys. Lett. {\bf 90}, 253507 (2007).

\bibitem{N2} E. H. Hwang, S. Adam and S. Das Sarma, Phys. Rev. Lett. {\bf 98}, 186806 (2007); S. Adam and S. Das Sarma, Solid State Commun. {\bf 146} 356 (2008). 


\bibitem{Raman} A.C. Ferrari \textit{et al}.,
  Phys. Rev. Lett. \textbf{97}, 187401 (2006). 

 

\bibitem{chen_phonon} J. H. Chen \textit{et al}., Nat. Nanotech. {\bf 3}, 206 
  (2008); S. V. Morozov, 
  K. S. Novoselov \textit{et al}., Phys. Rev. Lett. {\bf 100}, 016602 (2008);
  E. H. Hwang and S. Das Sarma,  Phys. Rev. B {\bf 77}, 115449 (2008)

\bibitem{tsephonon} W.-K. Tse and S. Das Sarma, Phys. Rev. Lett. {\bf
    99}, 236802 (2007). 
\bibitem{dassarma_ee} S. Das Sarma, E. H. Hwang, and W. K. Tse,
  Phys. Rev. B {\bf 75}, 121406 (2007). 
\bibitem{Ando} T. Ando, J. Phys. Soc. Jpn. \textbf{75}, 124701 (2006).
\bibitem{Polar} E. H. Hwang and S. Das Sarma, Phys. Rev. B {\bf 75},
  205418 (2007).  
\bibitem{jalabert}R. Jalabert and S. Das Sarma, Phys. Rev. B {\bf 40},
  9723 (1989). 
\bibitem{hwang_ee} E. H. Hwang and S. Das Sarma, Phys. Rev. B {\bf 77},
  081412(R) (2008); E. H. Hwang, B. Y. K. Hu, and S. Das Sarma,
  Phys. Rev. B {\bf 76}, 115434 (2007);  Physica E {\bf 40}, 1653 (2008).


\bibitem{dawlaty}J. M. Dawlaty \textit{et al}.,  Appl. Phys. Lett. {\bf 92}, 042116 (2008).


\bibitem{ltheta}  A. Palevski \textit{et al}., Phys. Rev. Lett {\bf
    62}, 1776 (1989);  T. Sakamoto \textit{et al}., Appl. Phys. Lett. {\bf 76}, 2618 (2000). 

\end{thebibliography}
\end{document}